\documentclass[12pt,a4paper]{article}
\usepackage{graphicx}
\usepackage{times}
\title{\bf Optical monitoring of Active Galactic Nuclei from ARIES}

\author{Gopal-Krishna$^1$\thanks{gopaltani@gmail.com} \ and Paul J. Wiita$^2$ \\
\vspace{1cm}\\
\normalsize $^1$ Centre for Excellence in Basic Sciences, Univ.\ of Mumbai Campus
(Kalina), Mumbai-400098, India\\ 
\normalsize $^2$ Department of Physics, The College of New Jersey, Ewing, NJ 08628-0718, USA \\
}
\date{\mbox{}}
\begin{document}
\maketitle
\pagestyle{empty}
%
%
\def\bull{\vrule height .9ex width .8ex depth -.1ex}
\makeatletter
\def\ps@plain{\let\@mkboth\gobbletwo
\def\@oddhead{}\def\@oddfoot{\hfil\scriptsize\bull\quad
"First Belgo-Indian Network for Astronomy \& astrophysics (BINA) workshop'', held in Nainital (India), 
15-18 November 2016 \quad\bull}%
\def\@evenhead{}\let\@evenfoot\@oddfoot}
\makeatother
%
%
\def\beginrefer{\section*{References}%
\begin{quotation}\mbox{}\par}
\def\refer#1\par{{\setlength{\parindent}{-\leftmargin}\indent#1\par}}
\def\endrefer{\end{quotation}}
%
%
{\noindent\small{\bf Abstract:} 

This overview provides a historical perspective highlighting the pioneering role which the fairly modest 
observational facilities of ARIES have played since the 1990s in systematically characterizing the 
optical variability on hour-like time scale (intra-night optical variability, or INOV) of several major 
types of high-luminosity Active Galactic Nuclei (AGN). Such information was previously available only for 
blazars. Similar studies have since been initiated in at least a dozen countries, giving a boost to AGN 
variability research. 

Our work has, in particular, provided strong indication that mild INOV occurs in radio-quiet QSOs 
(amplitude up to $\sim 3-5\%$ and duty cycle $\sim 10\%$) and, moreover, has demonstrated that similarly 
mild INOV is exhibited even by the vast majority of radio-loud quasars which possess powerful relativistic 
jets (even including many that are beamed towards us). The solitary outliers are blazars, the tiny 
strongly polarized subset of powerful AGN, which frequently exhibit a pronounced INOV. Among the blazars, 
BL Lac objects often show a bluer-when-brighter chromatic behavior, while the flat spectrum radio quasars 
seem not to. Quantifying any differences of INOV among the major subclasses of non-blazar type AGNs will 
require dedicated monitoring programs using $2-3$ metre class telescopes. 

\section{The groundwork}
The epochal discovery of quasars (Schmidt 1963) was quickly 
followed by the intriguing announcement of optical microvariability of the quasar 3C 48 by $\sim$4\% over 
just 15 minutes (Mathews \& Sandage 1963). In the following two decades, similar events of rapid optical 
variability were reported for a few blazars, the most active subset of Active Galactic Nuclei (AGN), as 
recounted in Jang \& Miller (1997). In hindsight, all these claims of blazar variability seem plausible, 
although they evoked persistent skepticism until the late 1980s when CCD cameras came to be deployed for 
intra-night optical monitoring of AGN, allowing simultaneous recording of many (comparison) stars on the 
same CCD chip. The advantage then is that they all are subjected to essentially the same air mass, weather 
and instrumental conditions. Hence, any variations due to these factors are effectively cancelled
 out simply by performing differential photometry between the AGN and stars in the same field-of-view. 
 This mode of using the CCD camera as a multi-star photometer routinely yielded reliable, sensitivity 
 enhanced ``differential light curves" (DLCs) of the monitored compact AGN relative to several comparison 
 stars (Miller et al.\ 1989; Carini et al.\ 1990; Wagner et al.\ 1990; Quirrenbach et al.\ 1991; Carini \& Miller 1992; Noble 
 et al.\ 1997). Such AGN variability was termed optical microvariability and later as, intra-night optical 
 variability (``INOV", see, Gopal-Krishna et al.\ 2003). It is somewhat amusing that the radio equivalent 
 of INOV of blazars had already been established  (Witzel et al.\ 1986; Heeschen 
 et al.\ 1987), though in many cases it could be largely due to propagation effects (e.g.,
 Shapirovskaya 1978; Rickett et al.\ 2001; Jauncey \& Macquart 2001;  Liu et al.\ 2011; de Bruyn \& Macquart 
 2015; Gopal-Krishna \& Subramanian 1991; also, Vedantham et al.\ 2017; but see Wagner et al. 1996; Krichbaum 
 et al.\ 2002; Gaba{\'n}yi et al.\ 2007; Wagner et al.\ 2008).
 
During the first few years, INOV searches remained focused on prominent BL Lacs, flat-spectrum radio-loud 
quasars (FSRQs) and Seyferts (e.g., Jang \& Miller 1995; see Gopal-Krishna et al.\ 
1995 for a summary). The first step towards diversifying to other major classes of high luminosity AGN 
was taken by one of us (G-K) in 1991, by taking intra-night `snapshots' of two radio-quiet QSOs (RQQSOs or RQQs) 
using the New Technology Telescope (NTT). Although neither RQQSO showed INOV (Gopal-Krishna et al.\ 1993a), 
we felt motivated to continue this program using the just commissioned 2.34-m Vainu Bappu Telescope (VBT) 
of the Indian Institute of Astrophysics (IIA). The resulting first paper reported a negative search 
for INOV in 5 high-luminosity RQQs (M$_B < - 25$), down to a 2--3\% detection threshold (Gopal-Krishna 
et al.\ 1993b). This study, coinciding with G-K's sabbatical visit to the Space Telescope Science 
Institute, was in fact the first VBT publication in an international journal (Rao 1995). Our next paper 
reported densely sampled intranight DLCs of 6 RQQs and one ``radio-intermediate quasar" (Gopal-Krishna 
et al.\ 1995). Among these RQQSOs, strong hints of INOV (amplitude $\psi \sim 5$\%) were seen in the DLCs 
of PG 0946$+$301 ($z = 1.216, M_V = -28.6$)  and also for PG 1630$+$377 ($z = 1.466, M_V = -29.5$). 
Additional examples of INOV of RQQs are reported in de Diego et al.\ (1998). The possibility had been mentioned 
that such INOV could be associated with transonic flows and shocks within the accretion disk (Chakrabarti 
\& Wiita 1992). Alternative accretion disk origins for INOV include an array of hotspots (e.g., Mangalam 
\& Wiita 1993), disk oscillations (Nowak \& Wagoner 1991, 1992), plasma instabilities (Krishan \& Wiita 
1994). Some other proposals invoke a clumpy torus (Nenkova et al.\ 2008) or reprocessing of coronal X-ray 
flares in the accretion disk (e.g., Rokaki et al.\ 1993; Merloni \& Fabian 2001); further possibilities 
are considered in Wiita (1996) and Raginski \& Laor (2016).

In an independent program, Jang \& Miller (1997) reported hour-scale micro-variability for two RQQSOs 
whose luminosities are close to the borderline between QSOs and Seyfert galaxies (cf.\ Miller et al.\ 
1990). Based on admittedly small samples, both they and we were led to conclude that INOV is distinctly 
rarer and milder among RQQSOs, as compared to blazars, establishing the inevitability of jet (vis a vis 
accretion disk) origin for a pronounced INOV. For the jet dominated sources, the strong relativistic 
enhancements of small fluctuations arising through turbulence (e.g., Marscher \& Travis1991; Romero et al.\ 2000; 
Goyal et al.\ 2012; Marscher 2014; Calafut \& Wiita 2015; Pollack et al.\ 2016), or ultra-relativistic 
mini-jets (Giannios et al.\ 2009, 2010; also, Singal \& Gopal-Krishna 1985) certainly dominate the 
short-term variability in blazars across all bands. The geometric explanations for rapid blazar variability 
include the `lighthouse effect' (Camenzind \& Krockenberger 1992) and the `swinging jet' scenario 
(Gopal-Krishna \& Wiita\ 1992; also, Bachev et al.\ 2012).

\section{The transition phase from VBT (IIA) to ST (UPSO) [1996 - 99]} 

The transition from the VBT to our new ``workhorse", the 1.04-m Zeiss telescope (the Sampurnand Telescope, 
hereafter ST) of the Uttar Pradesh State Observtory (UPSO, later ARIES), was a watershed. 
An update on our INOV program, particularly the INOV patterns of RQQs vs RLQs, is provided in 
Gopal-Krishna et al.\ (2000), together with the prevailing international setting which was still in a 
confused state (see, e.g., Jang \& Miller 1995, 1997; de Diego et al.\ 1998; Rabbette et al.\ 1998; 
Romero et al.\ 1999). As discussed below, this was partly due to the tendency to lump together diverse  
types of  radio-loud AGN (see, e.g., Helfand 2001; Carini et al.\ 2007; Ram{\'i}rez et al.\ 2009). Thus, a systematic 
characterization of the INOV of major classes of high-luminosity AGN became our focus.

Using the ST, we launched in 1998 a program of R-band intranight monitoring of 7 sets of bright (m$_B \sim 
16$) AGNs, each set falling in a narrow redshift bin within the overall range $z = 0.17$ to $2.2$. Each set 
consisted of a RQQ, a BL Lac (except in the highest $z$ bin), a radio lobe-dominated quasar (LDQ), and a 
core-dominated quasar (CDQ). The four major AGN classes in this sample are thus matched in the $z - M_B$ 
plane. We monitored each AGN for 4 to 8 hours per night, each on at least 3 nights, taking $\sim 5$ exposures 
per hour. This program took 113 nights (720 hours) during 1998--2002. All 7 RQQs are not only very 
luminous ($-24.3 > M_B > -29.8$) but are also genuinely radio-quiet, with $R < 1$, where $R$ is the 
rest-frame ratio of 5 GHz to 440 nm flux densities 
(Kellermann et al.\ 1994). Normally, intranight variations of 1-2\% could be detected in our DLCs.

The first publication from this UPSO program reported INOV of the RQQs B1029$+$329 ($M_B = -26.7, 
z = 0.560$) and B1252$+$0200 ($M_B = -24.8, z = 0.345$), albeit the INOV amplitude ($\psi$, defined in 
Romero et al.\ 
1999), was $< 3$\% for all 7 RQQs monitored, unlike the BL Lacs which overshot this limit in $\sim 50\%$
of the sessions (Gopal-Krishna et al.\ 2003; see, also Stalin et al.\ 2004a; Sagar et al.\ 2004; Carini 
et al.\ 2007; Gopal-Krishna et al.\ 2011). Thus, our data provided, for the first time, an estimate 
of INOV duty cycle (DC) for BL Lacs in different ranges of $\psi$. We further showed that even a violently 
intranight variable, like the BL Lac AO 0235$+$164, would exhibit a very low-level INOV (reminiscent of RQQs) 
were its jet misaligned from the line of sight by just 10-15$^{\circ}$, thanks to the drastic relativistic
flux de-boosting and time stretching (Gopal-Krishna et al.\ 2003; Stalin et al.\ 
2004a; also, Czerny et al.\ 2007).  INOV detections for RQQs were also reported from other observatories 
(e.g., Gupta \& Joshi 2005; Stalin et al.\ 2005; Ramirez et al.\ 2009; Polednikova et al.\ 2016).

The next step of our program was to compare the INOV characteristics of two AGN classes which appear 
strongly relativistically beamed, CDQs and BL Lacs. For this, matched samples of 5 CDQs and 6 BL Lacs 
were monitored in 46 sessions, each lasting $\sim 6.5$ hours (Sagar et al.\ 2004). This was the first
clear demonstration that the presence of even a relativistically beamed radio core is no guarantee of strong 
(blazar-like) INOV and that a high optical polarization is the critical factor, as amply confirmed
by Goyal et al.\ (2012; see below). 

Another novel step in our program was to systematically compare the INOV properties of RQQs and radio 
lobe-dominated quasars (LDQs). For this, 7 optically bright and high luminosity RQQs and 7 LDQs matched 
in the $z - M_B$  plane, were monitored in 61 sessions of $\sim 6$ hour duration. Once again, contrary to 
the folklore that INOV correlates with radio loudness, 
we found that the INOV characteristics ($\psi$, DC) of LDQs compare well in mildness with those of RQQs
(Stalin et al.\ 2004a,b).  
Since LDQs are universally believed to have powerful relativistic jets, our INOV result meant that RQQs need not be devoid of 
relativistic jets, either. This idea, in fact, accords well with the jet-disk symbiosis hypothesis (e.g., 
Falcke et al.\ 1995). More direct support has come from the VLBI detections of radio cores in many RQQs 
(e.g., Blundell \& Beasley 1998; Caccianiga et al.\ 2001; Ulvestad et al.\ 2005; Falcke et al.\ 1996; 
Leipski et al.\ 2006; Herrera Ruiz et al.\ 2016) and from radio flux variability (Barvainis et al.\ 2005; 
Wang et al.\ 2006) and detection of 
extended radio lobes (Kellermann et al.\ 1994; Blundell \& Rawlings 2001). Likewise, from a detailed 
analysis of the radio--far-IR correlation, White et al.\ (2017) have gathered credible evidence for 
AGN dominated radio flux of many RQQs. Note, however, that Raginski \& Laor (2016) have argued that the origin 
of the VLBI cores in RQQs needs not invoke nonthermal relativistic jets (see also Ishibashi \& Courvoisier 2011; Steenbrugge et al. 2011). 
Reverting to the compact jet 
scenario, the possibility has been advanced that RQQs may arise from an inverse Compton quenching of the 
jets in a majority of QSOs, before reaching the physical scale probed by radio emission (e.g., Brown 1990;
Barvainis et al.\ 2005; also, Xu et al.\ 1999). A possible signature of such quenching is the hard tail 
seen in the X-ray spectrum of several RQQs above $\sim 5$ keV (George et al.\ 2000). Note that detection of 
such emission is strongly disfavored because of the very narrow pattern of the inverse Compton boosting of 
external (e.g., broad-line region) photons by the relativistic jet (see, Dermer 1995). Thus, even though 
the jet scenario for RQQs does not discount starbursts  playing a major role in {\it long-term} optical 
variability (Terlevich, Melnick \& Moles 1987), it does highlight the relevance of INOV studies to 
the outstanding issue of the QSO radio dichotomy, reviewed, e.g., by Begelman et al.\ (1984), Urry \& 
Padovani (1995) and Antonucci (2012).

To recapitulate, it has come to be appreciated that radio loudness of a quasar (even if due to a jet
relativistically beamed towards us) is not a sufficient condition for detecting a pronounced INOV, as 
displayed by blazars. {\it There is no one-to-one mapping of the radio dichotomy of quasars to their INOV 
properties}. Remarkably, a similar conclusion has been reached from {\it radio} continuum variability on 
month-like time scales (Barvainis et al.\ 2005). We also note that the determination of INOV amplitude and 
duty cycle in our UPSO program, for representative sets of RQQs (DC $\sim$ 17\%), LDQs ($\sim$12\%), CDQs 
($\sim$20\%) and BL Lacs ($\sim$72\%), well matched in the $z - M_B$ plane, was a first time endeavour,
with the added advantage that the same instrumental set up and analysis procedure were used (and a typical 
INOV detection threshold of 1--2\% achieved). Additional support to the above INOV characterization for 
the 4 most prominent classes of high luminosity AGN has come by harnessing an enlarged INOV database 
through new observations and quality data mining from the literature (see below and Table 1). Here it is relevant to 
underline that, again somewhat unexpectedly, RQQs and CDQs show similar optical variability  
on week/month-like time scales (e.g., Bauer et al.\ 2009; Gaskell et al.\ 2006).

Early on, we pointed out that the photometric error given by the standard routine in IRAF is underestimated by a factor $\eta \simeq 1.5$ and ignoring this in statistical tests can often lead to spurious variability claims (Gopal-Krishna et al.\ 1995). This correction factor has now become a standard part of the $C-$test and $F-$test (Goyal et al.\ 2013b and references therein). 

\section{Continuation of the INOV program in the ARIES era}

Subsequent to the metamorphosis of UPSO into ARIES, our INOV program using the ST was pursued with an
expanded scope by including other important classes of AGN whose INOV characteristics were largely unknown. 
These classes are: Radio-Intermediate Quasars (RIQs) and TeV blazars (Goyal et al.\ 2010; Gopal-Krishna et 
al.\ 2011). Secondly, the role of optical polarization in INOV was more thoroughly assessed using a sample 
of 16 CDQs (Goyal et al.\ 2012). Lastly, our vast database of about 250 intranight DLCs covering 6 major classes
of AGN was uniformly re-analyzed by applying a more authentic statistical test, the $F$-test, newly 
emphasized by de Diego (2010). This yielded, for the first time, a uniform characterization of the INOV 
properties of 6 major classes of high-luminosity AGN, namely RQQs, RIQs, LDQs, LPCDQs, HPCDQs and 
TeV blazars (Goyal et al.\ 2013a; Table 1). The 1.04-m ST continued to be our workhorse even in this phase, 
although we did occasionally use the 2-m HCT of IIA and the 2-m Giravali telescope of IUCAA. 
The main results of this program are summarized below.

{\it (a) Radio-Intermediate-Quasars (RIQs)}: 
With a radio loudness parameter ($R$) in the range $\sim$ 3 to 100, RIQs are a link between RQQs and 
RLQs. The possibility of their being counterparts of RQQs with Doppler boosted radio jets has been
put forward (e.g., Miller et al.\ 1993; Falcke et al.\ 1996; Xu et al.\ 1999; Barvainis et al.\ 2005). 
Intranight R-band monitoring of 8 optically bright RIQs having flat or inverted radio spectra (hence 
jet dominated), was carried out on 25 nights during 2005--2009, yielding an estimate for INOV duty cycle, 
DC $\sim$ 10\% and $\psi \sim < 3$\%. This shows that in INOV properties, RIQs are close cousins of CDQs, 
rather than blazar-like (Goyal et al.\ 2010).

{\it (b) What holds the key to INOV: Polarization or relativistic beaming?}
This study was designed to probe the relationship of INOV with two key observational signatures of 
the blazar phenomenon, namely, optical polarization and radio core dominance. This point is relevant 
since, historically, a flat/inverted radio spectrum of a quasar, a marker of core dominance, has often
been deemed adequate for a blazar classification (e.g., Wills et al.\ 1992; Maraschi \& Tavecchio 
2003).  
We performed 44 nights of intranight monitoring (typical duration $\sim$ 5.7 hour) of a sample 
consisting of 12 low polarization core-dominated quasars (LPCDQs, $p < 2$\%) and 9 high polarization 
core-dominated quasars (HPCDQs, $p >$ 4\%). The two sets are otherwise matched in the $z - M_B$ plane 
and radio spectral index (Goyal et al.\ 2012). Whereas, for the HPCDQs, a large INOV ($\psi >$ 4\%) 
occurred on 11 out of 22 nights, the same was seen for the LPCDQs on just 1 out of 22 nights. This 
striking contrast was the first unambiguous demonstration that the physical link of INOV with optical 
polarization is far more fundamental than that with the relativistic beaming of the nuclear jet, as 
manifested by a dominant radio core, or a flat/inverted radio spectrum. Significantly, this tight 
correlation holds even if the optical polarization had been measured in a distant past, implying that that the 
propensity of a given FSRQ to exhibit (or, not exhibit) strong INOV is of a fairly stable nature 
(Goyal et al.\ 2012). These authors have also sketched a model for the correlation, in terms of a 
turbulent synchrotron plasma of the relativistically moving jet crossing standing shocks, just ahead 
of the jet's acceleration zone envisaged in the model by Marscher et al.\ (2008); also see Marscher 
(2014) and Pollack et al.\ (2016).

{\it (c) A search for INOV on sub-hour time scale (using TeV blazars)}:
Under this first INOV campaign dedicated to TeV detected blazars, we monitored 9 TeV blazars on 26 nights 
for an average duration of 5.3 hours, and then combined those data with similar INOV data gathered from 
the literature for another 13 TeV blazars (90 nights). It was thus shown (Gopal-Krishna et al.\ 2011) 
that the well known trend for the  low-peaked BL Lac objects (LBLs) to exhibit a distinctly stronger 
INOV, compared to high-peaked BL Lac objects (HBLs) (e.g., Heidt \& Wagner 1998; Romero et al.\ 2002) 
persists even when their TeV detected subsets alone are considered. Secondly, despite a dense and extensive
 intranight sampling, the DLCs of the 22 TeV blazars did not reveal even a single credible feature on a 
time-scale substantially shorter than 1 hour (see, also, Sagar et al.\ 1999). This is particularly 
noteworthy since extremely large bulk Lorentz factors $\Gamma > 50$ (and correspondingly stronger time 
compression) have been inferred from the brightness and rapid variability of TeV blazars, as their jets 
must be highly relativistic in order to avoid absorption of the TeV photons by co-spatial IR photons (e.g., 
Krawczynski et al.\ 2002; Aharonian et al.\ 2007; Begelman et al.\ 2008). Our 
search for sub-hour time scales was triggered by some claims (and counter-claims) in the literature on the 
issue of minute-scale optical variability of some blazars (e.g., as also noted in Romero et al.\ 2002). 
Here it is relevant to recall that even for EGRET blazars, the shortest time scale of INOV is seen to be 
$\sim 1$ hour (Romero et al.\ 2000). Note also that a search for sub-hour time scales in the radio light 
curves of some prominent intra-day variable blazars has also proved negative (Krichbaum et al.\ 2002).

The shortest variability time-scale is important for understanding the geometry of jets and 
the magnetic field, because it provides a possible maximum size of the variable region. The very short 
timescales probe the conditions very close to the centre. Our negative search demonstrates that, unlike 
at X-ray and $\gamma$-ray bands, optical variability on minute-like timescales is really rare, at 
least for amplitudes above $\sim 1$\%. Nonetheless, a 15-min time scale has been detected in the optical 
light curve of the BL Lac object S5 0716$+$714 (Sasada et al.\ 2008; Rani et al.\ 2010; also, Wagner et 
al.\ 1996). The size of the optically emitting region corresponding to such time scales begins to match the 
Schwarzschild radius of the central black hole in high luminosity quasars. 

\begin{table}
\caption{INOV Duty Cycles using the $F$-test}
\label{table1}
\small
\begin{center} 
\begin{tabular}{| l | c c c |}
\hline 
AGN type & No.\ of sessions & INOV DC & INOV DC ($\psi>3$\%) \\
\hline
Radio-quiet quasars (RQQs)                       &68       &10\%       & 6\%  \\
Radio-intermediate quasars (RIQs)               &31       &18\%        &11\%\\
Lobe-dominated quasars  (LDQs)                   &35       & 5\%        & 3\%\\
Low-polarization core dominated quasars (LPCDQs)   &  43      & 17\%       & 10\%\\
High-polarization core-dominated quasars (HPCDQs)  &   31      & 43\%      & 38\%\\
TeV detected blazars (TeVBLs)                     &85     &  45\%     &  45\%\\
\hline 
\end{tabular} 
\end{center} 
\end{table} 

{\it (d) Improved characterization of intranight optical variability of 6 prominent AGN classes (using 
the $F$-test)}:  Until about 2010, the statistical test commonly used for checking the presence of INOV 
in DLCs was the so called, $C$-test, introduced by Jang \& Miller (1997). However, de Diego (2010) 
questioned its validity on the ground that the $C$-statistic does not have a Gaussian description 
(as was commonly assumed). Instead, he advocated the use of $F$-test which compares the observed to 
expected variances (see, also Villforth et al.\ 2010). We therefore decided to re-evaluate the INOV 
characteristics of 6 major classes of powerful AGN, by applying the $F$-test to their existing $\sim~300$ 
DLCs based on the intranight monitoring with the ST. The resulting estimates of the INOV DC are given in 
Table 1 which also lists the DC values for INOV amplitude $\psi > 3\%$ (Goyal et al.\ 2013a). The motivation 
for this work was the need to establish a uniform benchmark for future INOV studies. 

{\it (e) Multi-week, simultaneous multi-colour intranight monitoring of prominent BL Lac objects}:
Multi-colour intranight monitoring of the prominent blazar S5 0716$+$714 was carried out in 1994 
(21 nights) and 1996 (11 nights) using ST and VBT and of BL Lacertae in 2001 for 5 nights (Sagar et al.\ 1999; 
Stalin et al.\ 2006). While both blazars showed strong INOV in these observations they also revealed 
a contrasting INOV behaviour, on intranight and possibly also on internight scale. While BL Lac became 
bluer when brighter, no chromatic trend was seen for the variations of S5 0716$+$714. Once again, 
no sub-hour variations were detected even in these extensive INOV observations (37 nights) of two of the 
most active AGN.
 
{\it (f)  Using INOV to search for the elusive radio-quiet BL Lacs}:
Having consistently observed that an INOV amplitude $\psi > 5$\% flashes a BL Lac,  we launched in 
2012 a program of intranight monitoring of a carefully selected sample of the enigmatic and rare 
subclass of quasars, called `radio-quiet weak-line QSOs' (RQWLQs). It 
has been speculated for many years that samples of RQWLQs might harbor some members of the long sought, 
elusive minuscule population of {\it radio-quiet} BL Lacs, in analogy to quasars, most of which are in fact
radio-quiet (e.g., Londish et al.\ 2002; Collinge et al.\ 2005; Stocke 2001; Shemmer et al.\ 2006). 
Since WLQs are mostly rather faint (m$_V > 18-19$m), this first-of-its-kind INOV project had to remain on 
hold until the installation of the 1.3-m telescope (`Devasthal Fast Optical Telescope': DFOT), our new 
workhorse. 

Results from this ongoing program are presented in five papers (Gopal-Krishna et al.\ 2013; Chand et al.\ 2014;
Kumar et al.\ 2015, 2016, 2017). These are based on intranight monitoring of 33 RQWLQs in 60 sessions 
lasting a minimum of 3 to 4 hours. Very briefly, two of the RQWLQs have exhibited episodes of strong INOV and 
another two have shown night-to-night variability, making them the best available candidates for the 
elusive radio-quiet BL Lacs. They are being further probed by optical polarimetric and continuum monitoring.
Besides the Indian telescopes, the 2-m telescope in Haute-Provence (France) has also contributed to this 
project.

\section{Major international collaborations}

Photometry of the blazars Mrk 421 and 3C 454.3 taken in 2009--2010 at ARIES using the 1.04 m ST was 
studied along with X-ray data from {\it MAXI} 
and $\gamma$-ray  measurements from {\it Fermi} 
(Gaur et al.\ 2012a).  Both objects were in bright states then, and genuine INOV was detected for both objects.
For Mrk 421 the X-ray fluctuations appeared to lead the optical ones by about 10 days. On the other hand, 
for 3C 454.3 the X-ray flux was not correlated with the optical (or $\gamma$) fluxes, but the $\gamma$-ray 
changes apparently led the optical ones by around 4 days (Gaur et al.\ 2012a).

Ten low-energy peaked blazars were observed at ARIES and Mt.\ Abu between 2008--2009 for an average of 
four nights each and a duty cycle for their INOV was found to be about 52\%; this is in accord with other 
measurements if the relatively short monitoring durations of $\sim$4 hours each are taken into account 
(Rani et al.\ 2011a).  In 2009--2010 a substantial number, 17, of radio-quiet broad absorption line 
quasars (BALQSOs) were monitored at ARIES and another two were monitored with the Himalayan Chandra 
Telescope in Ladakh, for $\sim$4 hours each (Joshi et al.\ 2011). The majority of these BALQSOs did not 
exhibit clear INOV (only 2 of 19 did) though several others showed hints of changes. Hence radio-quiet 
BALQSOs seem to show roughly the same low duty cycle for INOV that is displayed by normal RQQSOs (Joshi 
et al.\ 2011).


Over the past several years, Alok Gupta, his students, and others at ARIES have collaborated closely 
with a group of astronomers in Bulgaria to study INOV as well as longer term variability of many AGN, 
with a focus on blazars.  This research has typically combined roughly equal amounts of photometric 
data taken with the 1.04 m ST and/or with the 1.3 m DFOT, with observations from several  
telescopes in Bulgaria, Greece, and occasionally Serbia.  This set up can provide a longer continuous 
coverage of a given target on some nights, but more typically provides LCs on nights when only one of the 
observatories can successfully observe it. 

Optical flux and color variations of 12 low-to-intermediate energy peaked blazars were studied for 
short-term (day-to-month) variations  in 2008--2009 (Rani et al.\ 2010). Over those timescales, 92\% 
showed significant flux variability and 33\% showed clear color variations.  The six BL Lac objects 
usually became bluer-when-brighter, while the six FSRQs displayed a weak opposite trend (Rani et al.\ 
2010). The low energy (radio to optical) SEDs of 10 members of the above sample also were measured during 
2008--2009 (Rani et al.\ 2011b), usually at multiple epochs.  These could be well reproduced by a 
synchrotron model involving a log-parabolic distribution of electron energies and the variations in SEDs 
could be primarily attributed to changing jet Doppler factors (Rani et al.\ 2011b).

Multi-band optical observations of several blazars, including 3C 454.3, 3C 279 and S5 0716$+$714, all 
of which have been found to exhibit high INOV duty cycles, have frequently, but not consistently, shown 
bluer-when-brighter trends, particularly in overall active states (Agarwal et al.\ 2015; 2016). This 
type of chromatic change can be explained if the increasing synchrotron flux is related to a hardening 
of the underlying non-thermal electron spectrum and enhanced efficiency of particle acceleration.  
On the other hand, the high-energy peaked blazars 1ES 1959$+$650 and 1ES 2344$+$514 showed no genuine 
INOV over the course of observations covering 24 and 19 nights, respectively, in agreement with earlier 
studies suggesting that low-energy peaked blazars are more variable in optical bands (Gaur et al.\ 2012b
and references therein).  

BL Lac was in an active phase during 2010--2013 and ARIES observations were a key part of long term 
multi-band optical/NIR monitoring (Gaur et al.\ 2015a; see Fig.\ 1) that were combined with radio 
monitoring conducted in Finland and Ukraine at several frequencies (Gaur et al.\ 2015b). During this 
period the cm-fluxes were delayed with respect to the optical by $\sim$250 days, but no typical 
variability timescales were found for either, implying an intrinsic origin for the radio variability. 
Additional multi-band monitoring of BL Lacertae was conducted on 13 nights in late 2014; it showed 
INOV on several of these nights and also displayed the bluer-when-brighter trend (Agarwal \& Gupta 2015). 
Between 2012 and 2014, extensive observations of 3 TeV blazars, PKS 1510$-$089, PG 1553$+$113 and 
Mrk 501 were performed in the B, V, R and I bands at ARIES and in Bulgaria, Greece and Serbia.  
All three blazars remained active throughout the campaign, but no significant evidence was seen 
for spectral changes being correlated with the brightness of the source (Gupta et al.\ 2016).

\begin{figure}[h]
\centering
\includegraphics[width=17.3cm]{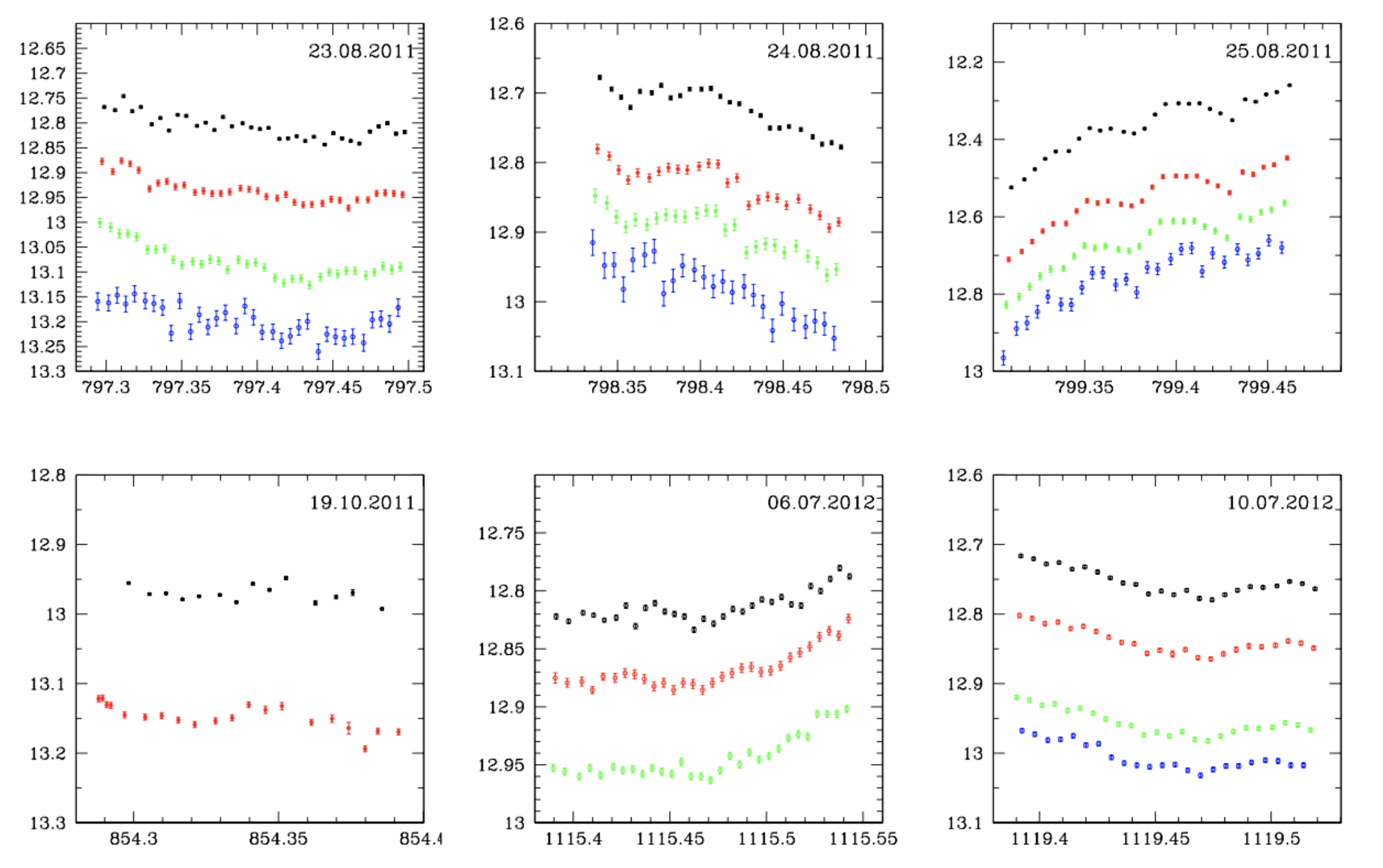}
\caption{Multi-band intranight LCs of BL Lacertae in 2011 and 2012; B (blue), V (green), R (red) and I 
(black) bands, reproduced with permission from part of Figure 2 in Gaur et al.\ (2015a), MNRAS, 452, 4263. 
The X-axes are JD (2455000$+$) and the Y-axes are the magnitudes in each panel, with the B, V and I bands 
shifted by arbitrary offsets with respect to the best calibrated R-band LC. \label{fig_1}}
\end{figure}


Observations from ARIES have also contributed to several major ``Whole Earth Blazar Telescope" (WEBT) 
campaigns that attempted to obtain essentially continuous LCs of particularly active objects for several
consecutive days, by combining optical observations made with many ground based telescopes at different 
longitudes, so that the source always remains accessible to at least one telescope. In most cases, these 
WEBT campaigns are accompanied by simultaneous radio and X-ray, and sometimes with IR, UV and $\gamma$-ray,
monitoring as well. Such coordinated multi-band observations of AGN yield instantaneous SEDs and their 
change with time. These measurements, particularly the temporal lags between fluctuations in different 
bands, provide critical information on the nature and location of the processes responsible for the 
acceleration of electrons to ultra-relativistic energies. WEBT campaigns to which ARIES contributed are 
those during 2007--2008 on 3C 454.3 (Raiteri et al.\ 2008), 3C 66A  (B{\"o}ttcher et al.\ 2009) and 
BL Lacertae (Raiteri et al.\ 2009), as well as one during February 2009 on S5 0716$+$714 (Bhatta et al.\  
2013).  

ARIES led a similar multi-site, multi-band, campaign on S5 0716$+$714 for a week during December 2009 
(Gupta et al.\ 2012):  the key results were that large variations, with similar timescales, were seen 
in the radio and optical bands, but not in X-rays. Analyses of these data indicate that the radio 
fluctuations were dominated by interstellar scintillations, the optical changes arose from intrinsic 
synchrotron emission, while the X-rays originated from inverse Compton scatterings. Together with 
nearly simultaneous $\gamma$-ray observations from {\it Fermi}, these data imply that the relativistic 
jet had a Doppler factor between 12 and 26, which is atypically large for a BL Lac (Gupta et al.\ 2012).

\section{Epilogue}
 The phenomenon of AGN Intra-Night Optical Variability (INOV) earned wide acceptance and traction in the 
 late 1980s, with the use of CCD cameras as multi-star photometers. However, this area of research was 
 then pursued at just a couple of observatories in the USA and was basically confined to blazars, a tiny subset
 of Active Galactic Nuclei. In 
 the early 1990s, its  major diversification by the present authors to encompass several other prominent 
 classes of powerful AGN, and its ensuing rapid growth and globalization has resulted in greatly enlarging 
 the footprints of the INOV research, to at least a dozen countries spread across four continents. 
 India's pioneering role in this diversification process has involved all major Indian optical 
 observatories, leading to 101 refereed publications in international journals and receiving over 1900 
 citations. ARIES (formerly UPSO) has been at the forefront of this activity which is now poised for 
 further expansion, taking advantage of its new telescopes at Devsthal, the 1.3-m DFOT and the just 
 commissioned 3.6-m DOT.

%
%
\section*{Acknowledgements}
The seeding of this research activity traces back to the present author G-K's visit to the ESO at La Silla in 
the late 1991, under the ESO Senior Visitor Program. For this, he is grateful to Prof.\ H.\ van der Laan 
and also wishes to express his thanks to the ESO colleagues, in particular, Prof.\ Jorge Melnick, 
Dr.\ Bruno Altieri and Prof.\ Massimo Della Valle for support in various ways.  PJW is grateful for 
sabbatical hospitality at KIPAC, Stanford University, as well as tremendous hospitality at NCRA/TIFR, IIA 
and ARIES over the decades.  G-K is supported by a NASI
Platinum Jubilee Senior Scientist Fellowship.  
%
%
%

\footnotesize
\beginrefer

\refer Agarwal A. and Gupta A.~C. 2015, MNRAS, 450, 541  

\refer Agarwal A., Gupta A.~C., Bachev R., et al. 2015, MNRAS, 451, 3882  

\refer Agarwal A., Gupta A.~C., Bachev R., et al. 2016, MNRAS, 455, 680  

\refer Aharonian F., Akhperjanian A.~G., Bazer-Bachi A.~R., et al. 2007, ApJ, 664, L71  

\refer Antonucci R. 2012, A\&AT, 27, 557  

\refer Bachev R., Semkov E., Strigachev A., et al. 2012, MNRAS, 424, 2625  

\refer Barvainis R., Leh{\'a}r J., Birkinshaw M. ,et al. 2005, ApJ, 618, 108 

\refer Bauer A., Baltay C., Coppi P., et al.\ 2009, ApJ, 699, 1732

\refer Begelman M.~C., Blandford R.~D., and Rees M.~J.  1984, RvMP, 56, 255  

\refer Begelman M.~C., Fabian A.~C., and Rees M.~J. 2008, MNRAS, 384, L19  

\refer Bhatta G., Webb J.~R., Hollingsworth H., et al. 2013, A\&A, 558, 92 

\refer Blundell K.~M. and Beasley A.~J. 1998, MNRAS, 299, 165  

\refer Blundell K.~M. and Rawlings S. 2001, ApJ, 562, L5  

\refer B{\"o}ttcher M., Fultz K., Aller H.~D., et al. 2009, ApJ, 694, 174 

\refer Brown R. L., 1990, in Zensus J. A., Pearson T. J. eds, Parsec-scale Radio
                           Jets. Cambridge Univ. Press, Cambridge, p.\ 199         

\refer Caccianiga A., Marcha M., Thean A., and Dennett-Thorpe J. 2001, MNRAS, 328, 867 

\refer Calafut V. and Wiita P.~J. 2015, JApA, 36, 255  

\refer Camenzind M.\ and Krockenberger M. 1992,  A\&A, 255, 59  

\refer Carini M.~T., and Miller H.~R. 1992, ApJ, 385, 146  

\refer Carini M.~T., Miller H.~R., and Goodrich B.~D. 1990, AJ, 100, 347 

\refer Carini M.~T., Noble J.~C., Taylor R., and Culler R. 2007, AJ, 133, 303  

\refer Chakrabarti S.~K. and Wiita P.~J. 1992, ApJ, 387, 21L  

\refer Chand H., Kumar P., and Gopal-Krishna  2014, MNRAS, 441, 726  

\refer Collinge M.~J., Strauss M.~A., Hall P.~B., et al. 2005, AJ, 129, 2542  


\refer Czerny B., Siemiginowska A., Janiuk A., and Gupta, A.~C. 2007, MNRAS, 386, 1557  

\refer de Bruyn A. G. and Macquart J.-P. 2015, A\&A, 574, 125

\refer de Diego J.~A. 2010, AJ, 139, 1269  

\refer de Diego J.~A., Dultzin-Hacyan D., Benitez E., and Thompson K.~L. 1998  A\&A, 330, 419  

\refer Dermer C.~D. 1995, ApJ, 446, L63   

\refer Dondi L. and Ghisellini G. 1995, MNRAS, 273, 583  

\refer Falcke H., Malkan M.~.A., and Biermann P.~L. 1995, A\&A, 298, 375  

\refer Falcke H., Patnaik A.~R., and Sherwood W. 1996, ApJ, 473, L13  

\refer Gaba{\'n}yi K.~{\'E}., Marchili N., Krichbaum T.~P., et al. 2007, AN, 328, 863


\refer Gaskell C.~M., Benker A.~J., Campbell J.~S., et al. 2006, ASPC, 390, 41  

\refer Gaur H., Gupta A.~C., and Wiita P.~J. 2012, AJ, 143, 23  

\refer Gaur H., Gupta A.~C., Bachev R., et al. 2015a, MNRAS, 452, 4263  

\refer Gaur H., Gupta A.~C., Bachev R., et al. 2015b, A\&A, 582A, 103  

\refer Gaur H., Gupta A.~C., Strigachev A., et al. 2012b, MNRAS, 420, 3147
  
\refer George I.~M., Turner T.~J., Yaqoob T., et al. 2000, ApJ, 531, 52  

\refer Giannios D., Uzdensky D.~A., and Begelman M.~C. 2009, MNRAS, 395, L29  

\refer Giannios D., Uzdensky D.~A., and Begelman M.~C. 2010, MNRAS, 402, 1649  

\refer Gopal-Krishna, and Subramanian K. 1991, Nature, 349, 766  

\refer Gopal-Krishna, and Wiita P.~J. 1992, A\&A, 259, 109  

\refer Gopal-Krishna, Goyal A., Joshi S.,  et al. 2011, MNRAS, 416, 101  

\refer Gopal-Krishna, Gupta A.~C., Sagar R. et al. 2000, MNRAS, 314, 815 

\refer Gopal-Krishna, Joshi R., and Chand H.  2013, MNRAS, 430, 1302

\refer Gopal-Krishna, Sagar R., and Wiita P.~J. 1993b, MNRAS, 262, 963  

\refer Gopal-Krishna, Sagar R., and Wiita P.~J. 1995, MNRAS, 274, 701 

\refer Gopal-Krishna, Stalin C.~S., Sagar, R., and Wiita P.~J. 2003, ApJ, 586, L25

\refer Gopal-Krishna, Wiita P.~J., and Altieri B. 1993a, A\&A, 271, 89  

\refer Goyal A., Gopal-Krishna, Joshi S., et al. 2010, MNRAS, 401, 2622  

\refer Goyal A., Gopal-Krishna, Wiita P.~J., et al. 2012, A\&A, 544, 37  

\refer Goyal A., Gopal-Krishna, Wiita P.~J., et al. 2013a, MNRAS, 435, 1300    

\refer Goyal A., Mhaskey M., Gopal-Krishna et al.\ 2013b, JApA, 34, 273


\refer Gupta A.~C., Agarwal A., Bhagwan J., et al. 2016, MNRAS, 458, 1127  

\refer Gupta A.~C., and Joshi, U. C. 2005, A\&A , 440, 855  

\refer Gupta A.~C., Krichbaum T.~P., Wiita P.~J., et al. 2012, MNRAS, 425, 1357  

\refer Heeschen D.~S., Krichbaum Th., Schalinski C.~J., and Witzel A. 1987, AJ, 94, 1493  


\refer Heidt J. and Wagner S.~J. 1998, A\&A, 329, 853  

\refer Helfand D.~J. 2001, PASP, 113, 1159

\refer Herrera Ruiz N., Middelberg E., Norris R.~P., and Maini A. 2016, A\&A, 598, L2  


\refer Ishibashi A. and Courvoisier T.~J.-L.  2011, A\&A, 525, A118

\refer Jang M. and Miller  H.~R. 1995,  ApJ, 452, 582 

\refer Jang M. and Miller, H.~R. 1997, AJ, 114, 565  
 
\refer Jauncey D.~L. and Macquart J.~P.  2001, A\&A, 370, L9  
 
\refer Joshi R., Chand H., Gupta A.~C., and Wiita P.~J. 2011, MNRAS, 412, 2717  



\refer Kellermann K.~I., Sramek R.~A., Schmidt M., Green R.~F., and Shaffer D.~B. 1994, AJ, 108, 1163 



\refer Krawczynski H., Coppi P.~S., and Aharonian F. 2002, MNRAS, 336, 721 

\refer Krichbaum T.~P., Kraus A., Fuhrmann L., Cim{\`o} G. and Witzel A. 2002, PASA, 19, 14  

\refer Krishan  V. and Wiita P.~J. 1994, ApJ, 423, 172  

\refer Kumar P., Chand H., and Gopal-Krishna 2016, MNRAS, 461, 666  

\refer Kumar, P., Gopal-Krishna, and Chand, H. 2015, MNRAS, 448, 1463  

\refer Kumar, P., Gopal-Krishna, Stalin, C. S., et al. 2017 (submitted to MNRAS)  

\refer Leipski C., Falcke H., Bennert N., and H{\"u}ttemeister S. 2006, A\&A, 455, 161  

\refer Liu Y., Fan J.~H., Wang H.~G., and Deng G.~G. 2011, JApA, 32, 173  

\refer Londish D., Croom S.~M., Boyle B.~J., et al. 2002, MNRAS 334,, 941  

\refer Mangalam A.~V. and Wiita P.~J. 1993, 406, 420  

\refer Maraschi, L. and Tavecchio, F. 2003, ApJ, 593, 667  

\refer Marscher A.~P. 2014, ApJ, 780, 87  

\refer Marscher  A.~P., Jorstad S.~G., D'Arcangelo F.~D., et al. 2008, Nature, 452, 966

\refer Marscher A.~P. and Travis J.~P. 1991 in Variability of Active Galactic Nuclei, ed. H.~R.\ Miller, P.~J.\ Wiita \& J.~C.\ Noble, Cambridge Univ.\ Press, Cambridge,  p.\ 153  

\refer Matthews T.~A. and Sandage A.~R. 1963, ApJ, 138, 30 

\refer Merloni A. and Fabian A.~C., 2001, MNRAS, 328, 958  


\refer Miller H.~R., Carini M.~T., and Goodrich B.~D. 1989,  Nature, 337, 627 

\refer Miller J.~S., Peacock J.~A., and Mead A.~R.~G. 1990, ApJ, 244, 207  

\refer Miller P., Rawlings S., and Saunders R. 1993, MNRAS, 263, 425  

\refer Nenkova M., Sirocky M.~M., Nikutta R., Ivezi{\'c} {\^Z}., and Elitzur M.  2008, ApJ, 685, 160  

\refer Noble J.~C., Carini M.~T., Miller H.~R., and Goodrich B. 1997, AJ, 113, 1995  

\refer Nowak M.~A. and Wagoner R.~V. 1991, ApJ, 378, 656  

\refer Nowak M.~A. and Wagoner R.~V. 1992, ApJ, 393, 697  

\refer Polednikova, J., Ederoclite, A., de Diego, J. A., et al. 2016, MNRAS, 460, 3950 

\refer Pollack M., Pauls D., and Wiita P.~J. 2016, ApJ, 820, 12  



\refer Quirrenbach A., Witzel A., Wagner S., et al. 1991, ApJ, 372, L71


\refer Rabbette M., McBreen B., Smith N., and Steel, S.  1998, A\&AS, 129, 445  

\refer Raginski I. and  Laor A. 2016, MNRAS, 459, 2082  

\refer Raiteri C.~M., Villata M., Capetti A., et al. 2009, A\&A, 507, 769  

\refer Raiteri C.~M., Villata M., Larionov, V.~M., et al. 2008, A\&A, 491, 755 

\refer Ram{\'i}rez A., de Diego J.~A., Dultzin D., and Gonz{\'a}lez-P{\'e}rez J.-N.  2009, 138, 991 

\refer Rao N.~K. 1995, BASI, 23, 351 

\refer Rani B., Gupta A.~C., Bachev R., et al. 2011b, MNRAS, 417, 1881

\refer Rani B., Gupta A.~C., Joshi U.~C., et al. 2011a, MNRAS, 413, 2157 

\refer Rani B., Gupta A.~C., Strigachev A., et al. 2010, MNRAS, 404, 1992 

\refer Rickett B.~J., Witzel A., Kraus A., Krichbaum T.~P., and Qian S.~J. 2001, ApJ, 550, 11 
 
\refer Rokaki E., Collin-Souffrin S., and Magnan C.  1993, A\&A, 272, 8  
 
\refer Romero G.~E., Cellone S.~A., and Combi J.~A. 1999, A\&AS, 135, 477  

\refer Romero G.~E., Cellone S.~A., and Combi J.~A. 2000, A\&A, 360, L47  

\refer Romero G.~E., Cellone S.~A., Combi J.~A., and Andruchow I. 2002, A\&A, 390, 431 

\refer Sagar R., Gopal-Krishna, Mohan V., et al. 1999, A\&AS, 134, 453 

\refer Sagar R., Stalin C.~S., Gopal-Krishna, and Wiita P.~J.  2004, MNRAS, 348, 176  

\refer Sasada M., Uemura M., Arai A.,  et al. 2008, PASJ, 60, L37  

\refer Schmidt M. 1963, Nature, 197, 1040  

\refer Shemmer O., Brandt W.~N., Schneider D.~P., et al. 2006, ApJ, 644, 86  

\refer Shapirovskaya, N. Y. 1978, AZh, 55, 953

\refer Singal A. and Gopal-Krishna  1985, MNRAS, 215, 383  

\refer Stalin C.~S., Gopal-Krishna, Sagar R., and Wiita P.~J. 2004a, MNRAS, 350, 175  

\refer Stalin C.~S., Gopal-Krishna, Sagar R., and Wiita P.~J. 2004b, JApJ, 25, 1  

\refer Stalin C.~S., Gopal-Krishna, Sagar R. et al. 2006, MNRAS, 366,1337  

\refer Stalin C.~S., Gupta A.~C., Gopal-Krishna, et al. 2005, MNRAS, 356, 607

\refer Steenbrugge K.~C., Jolley E.~J.~D., Kuncic Z., and Blundell K.~M.  2011, MNRAS, 413, 1735



\refer Terlevich, R., Melnick, J., \& Moles, M. 1987, in IAU Symp. 121,
Observational Evidence of Activity in Galaxies, ed.\ E.\ E.\ Khachikian,
K.\ J.\ Fricke, \& J.\ Melnick, Kluwer, Dordrecht, p.\ 499                   

\refer Ulvestad J.~S., Antonucci R.~R.~J., and Barvainis R.  2005, ApJ, 621, 123  

\refer Urry C.~M. and Padovani P. 1995, PASP, 107, 803  

\refer Vedantham H.~K., Readhead A.~C.~S., Hovatta T., et al., 2017, arXiv170206582  

\refer Villforth C., Koekemoer A.~M., and Grogin N.~A. 2010, ApJ, 723, 737  

\refer Wagner S.~J., Sanchez-Pons F., Quirrenbach A., and Witzel A.  1990, A\&A, 235, L1

\refer Wagner S.~J., Witzel A., Heidt J., et al. 1996, AJ, 111, 2187  


\refer Wang T.-G., Zhou H.-Y., Wang J.-X., Lu Y.-J. and Lu Y. 2006, ApJ, 645, 856  


\refer White, S.~V., Jarvis M.~J., Kalfountzou E., et al. 2017, MNRAS, 468, 217  

\refer Wills B., Wills D., Breger M., Antonucci R.~R.~J., and Barvainis R. 1992, ApJ, 398, 454

\refer Wiita P.~J. 1996, in Blazar Continuum Variability, ed.\ H.~R.\ Miller, J.~R.\ Webb, \& J.~C.\ Noble, ASPC, 110, 42  

\refer Witzel A., Heeschen D.~S., Schalinski C., and Krichbaum Th. 1986, MitAG, 65, 239 

\refer Xu C., Livio M., and Baum S. 1999, AJ, 118, 1169

\endrefer       

\enddocument